\documentclass[letterpaper,twocolumn,10pt]{article}
\usepackage{usenix2019_v3}
\usepackage{longtable}
\usepackage{tikz}
\usepackage{amsmath}

\usepackage{bm}

\begin{document}

\date{}

\title{\Large \bf Intelligent Replication Management for HDFS Using Reinforcement Learning}

\author{
{\rm Hyunsung Lee}\\
Sungkyunkwan University\\
hyunsung.lee@skku.edu
}

\maketitle

\begin{abstract}
Storage systems for cloud computing merge a large number of commodity computers into a single large storage pool. It provides high-performance storage over an unreliable, and dynamic network at a lower cost than purchasing and maintaining large mainframe. In this paper, we examine whether it is feasible to apply Reinforcement Learning(RL) to system domain problems. Our experiments show that the RL model is comparable, even outperform other heuristics for block management problem. However, our experiments are limited in terms of scalability and fidelity. Even though our formulation is not very practical,applying Reinforcement Learning to system domain could offer good alternatives to existing heuristics.
\end{abstract}

\section{Introduction}
Storage systems for cloud computing merge a large number of commodity computers into a single large storage pool. 
It provides high-performance storage over an unreliable, and dynamic network at a lower cost than purchasing and maintaining large mainframe. 
One of the most famous solutions to build such a system is Hadoop Distributed File System (HDFS).

HDFS stores data by dividing data into small and fixed sized blocks (generally 128MB except for the last block of a file). HDFS stores these blocks over nodes on the network. Blocks are replicated, or stored more than one nodes. Replication has two purposes. First, it is used to prevent data loss at node failures. If a node crashes, data replicated to other nodes which is alive now can be retrieved but otherwise not. Secondly, by having multiple replicas of a block, several data reads can occur in parallel. Moreover, a client will access one of the replicas which is close to her to faster access and lower overhead on the network.

HDFS uses a simple replication policy (i.e., have three replicas for each block) with considering data locality and ensuring data availability in the event of failures. 
One block is stored at a node with the same rack with the original block, and another one is stored at a node in a different rack.
This replication policy also enables some degree of load balancing by distributing read accesses across nodes.
However, HDFS default replication policy has two drawbacks. 
Some blocks are retrieved more than other blocks. 
These "hot" blocks deserve larger replication factors. 
On the other hands, there are "cold" blocks that are rarely read. To serve these requests better, Larger number of replication count than three, thus larger number of different nodes is desirable for hot blocks but cold blocks a have lower number of replication count. ~\cite{Cheng2012}
However, in HDFS, the number of reads does not affect replication factors.
Current HDFS does not consider disk characteristics such as throughput and latency. 
Therefore there will be a room for performance optimizations if a block write and erase are managed with knowledge of data access patterns

Unfortunately, replication policy that exploits block and storage characteristic requires expert knowledge and significant efforts. ~\cite{Ciritoglu2018}.
\section{Proposed Approach}
Recent successes of applying machine learning and reinforcement learning based approaches to system domains suggest that the idea of managing the replication of distributed data storage plausible. Particularly, Reinforcement Learning has become an active area in machine learning researches. It has a long history but gets great attention since it has been combined with Deep Learning techniques to achieve successes at various domains such as playing video games~\cite{mnih2013playing}, device placement for deep learning~\cite{mirhoseini2017device}, scheduling spark jobs~\cite{mao2018learning}, and others.
Upon these successes, We believe that Reinforcement Learning can be used to solve this problem without expert knowledge of data replication and networks by automatically learning from data.

We propose an algorithm to decide appropriate replication factors by observing data access pattern and network traffic. It means that we have to answer these two questions, “where to put the new replica” (in case of increment of the number of replicas), and “which replica from replicas should We remove?” (in case of decrement of the number of replicas). Our model has to support these feature, that, placing a replica to an appropriate node. Moreover, it also takes heterogeneous storages into account to select an appropriate node to place blocks.

\section{Design} \label{sec:design}
In this section, the design for block replication manager is presented. We formulate the problem and describe how to represent it in the framework of Reinforcement Learning.

\subsection{Background} \label{subsec:background}
We briefly review Reinforcement Learning techniques used in the paper.

\subsubsection{Reinforcement Learning}

\begin{figure}[h]
\centering
\includegraphics[width=0.45\textwidth]{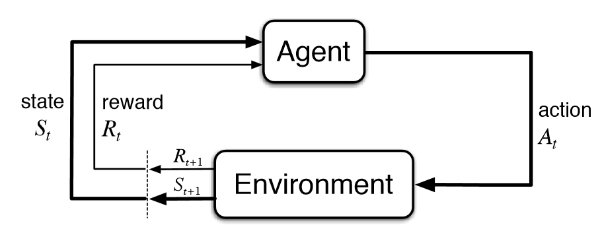}
\caption{General view of Reinforcement Learning.}
\end{figure}

Consider a situation where an agent interacts with an environment. At each time step $t$, the agent observes some state $s_t$, chooses an action $a_t$. Upon action $a_t$, the environment transits into state $s_{t+1}$ and emits reward $r_t$. The state transitions and rewards can be stochastic but are assumed to depend only on the action $a_t$. The agent interacts with the environment by observing state $s_t$, doing action $a_t$, and receiving $r_t$.

The goal of reinforcement learning is to teach the agent to choose action $a$ appropriately to maximize rewards it receives. Precisely, optimizing the agent is to make agent behave well to maximize cumulative discounted reward $\text{E}[\sum \gamma^t r_{t}]$ where $\gamma \in (0, 1]$ is a discounting factor to discount future rewards.

\subsubsection{Policy} \label{subsubsec:policy}
The agent chooses actions based on a policy, defined as a probability distribution over actions $\pi: s -> R^|A|\cap[0, 1]^|A|$. It is common to use function appoximators using neural network models. We refer to the function approximator of $pi$ with parameter $\theta$ as $\pi_{\theta}(s)$. We also used a neural network model to represent the policy in our design.

\subsection{modelling HDFS as view of RL}
\subsubsection{Environment Design} \label{subsubsec:env_design}
We consider a cluster with $M$ nodes, where nodes can store chunks of data, or blocks. 
There is a client that requests blocks. a client can directly access blocks not via master node. 
We assumed that all nodes are homogeneous, Throughput and response time of the nodes are identical.
Even though block reads and replication occur in a continuous manner(i.e., each block read request occur real-time, and We re-allocate block to other node or reduce the replication count of the block real-time), We designed it discrete manner. For example, data read occurs in given amount time $\delta t$ and We observe the block request pattern and decide which block to replicate, or move to another block or reduce a replication count of the block. Further, We do not consider effects of networking. All blocks can be accessed immediately(response time is zero), network congestion does not affect throughput, and there is no limit in the throughput in the network. In other words, a client can read any block she tries to read immediately. This simplification harms the fidelity of the simulation and environment. however, it still has the important aspects of block replication problem.

\subsubsection{State space}
We represent the state as lists of read counts of blocks that a node has of all nodes. In other words, The state is represented as a matrix $S$ of $M \times B$, where $M$ is the number of the nodes in the system, $B$ is the number of maximum blocks that a node can have. The element of the matrix $S_{i, j}$ is the $j$'th number of read counts at last time step. We sorted the read counts of the blocks decreasing order. We used 
\begin{itemize}
    \item \textbf{Node representation}
    We represent the state as lists of read counts of blocks that a node has of all nodes. In other words, The state is represented as a matrix $S$ of $M \times B$, where $M$ is the number of the nodes in the system, $B$ is the number of maximum blocks that a node can have. An element of the matrix $S_{i, j}$ is the $j$'th number of read counts at last time step. We sorted the read counts of the blocks decreasing order.
    \item \textbf{Block representation}
    Blocks are represented as a vector $v_b$of size $M$. $i$'th element of vector $v_b$ is one if block $b$ is in the node $i$, otherwise is zero.  representation of total nodes become $C$ by $M$, where $C$ is the number of distinct blocks in the system.
\end{itemize}

Finally, we flattened both matrices and concatenated to generate input to be fed to the RL model. total size of input is $M \times (B + C)$.

\subsubsection{Action space}
There are three kinds of actions. 
\begin{itemize}
  \item Copy a block $b$ in the node $m_i$ to $m_j$, increasing the replication count by 1. If a model tries to copy a block which has been reached maximum replication count, the action is ignored.
  \item Remove a block $b$ from the node $m$, decreasing the replication count by 1. If a model chooses to remove a block that has replication count 1, the action is ignored because zero replication is not allowed(data loss)
  \item Move a block $b$ in the node $m_i$ to $m_j$. If a model tries to copy a block which has been reached maximum replication count, the action is ignored. This action, actually, can be done using two actions mentioned above, but we designed it as distinct action because performing two actions in single timestep is not allowed.
\end{itemize}

Thus, Actions space has size of $M \times M \times 3$, to choose (1) a node from which we take a block, (2) a node to which a block is stored, (3) the number of action types.

Even though our action is to choose an reallocate block to the other node, we does not allow RL model, and baseline model to choose block directly. Instead, models and baselines choose a node, or from-node where a block to move, and a node(to-node) where a block will be stored. A block with a maximum read requsts in the from-node to be reallocated is chosen.

\subsubsection{Rewards}
We use the variance of the number of read request to each node as the objective to optimize. Formally, for each node $m$, the number of read request between $t$ and $t +\delta t$ $a_{m, t}$ can be calculated. 
We can then define the variance of read requests among nodes
$$
    \text{var}_m[a_{m, t}] = \frac{1}{M} \sum_{m}^M ({a_{m, t}} - \text{E}[a_{m, t}])^2
$$ where $\text{E}[a_{m, t}] = \frac{1}{M} \sum{a_{m, t}}$ is the mean of read requests for each node.

If read requests are distributed well between all nodes then the variance of the number of read requests among nodes will be small. Otherwise, the variance will be high.
We used $-\tau \times \text{var}_m[a_{m, t}]$ as reward of RL modelling. The model optimizes to minimize variance of job requests among nodes. $tau$ is the normalizing constant.

\section{Evaluation}
We perform primary evaluation of our model to answer following questions.

\begin{itemize}
    \item Is RL model able to learn a policy that minimizes variances of block read request among nodes?
    \item Does RL perform better than other heuristics?
\end{itemize}

\subsection{Methodology} \label{subsec:method}
\subsubsection{System} \label{subsubsec:system}
We implemented the system described in \ref{subsubsec:env_design}. Job access pattern follows a Zipf Distribution \cite{adamic2011complex}, meaning a few frequently accessed jobs take a most portion of job requests. Specifically, We created three Zipf distribution and sampled job requests from three of distributions, allowing three or more popular jobs exist in a timestep. Some times after, we create other Zipf distributions and sample jobs from new Zipf distributions(i.e., "Hot" blocks vary over time.). The number of block read requests are sampled from Poisson Distribution with mean of 200(i.e., there are 200 read requests in average at each timestep.).

A single node can have 120 jobs at maximum. Maximum replication count is set to be 5. 

Our environment ignores invalid action that the RL model can make, such as reducing replication count of a block to zero or copying a block that already reached maximum replication count. The RL policy makes action above, but those actions are ignored and have no effect at all.

The number of nodes vary 4 to 8, increasing by 2. The number of blocks can be either 128, or 256. 

\subsubsection{RL model} \label{subsubsec:rl_model}
As mentioned in \ref{subsubsec:policy}, we used neural network with single hidden layer of width 128 to model policy. We used Proximal Policy Optimization \cite{schulman2017proximal} with learning rate 0.001 to train RL models. We trained the model for 500,000 timesteps.

\subsubsection{comparables} \label{subsubsec:baseline}
We tested two RL models. First RL model is that we have three actions mentioned above as possible actions. We call this RL model RL-e model rest of the paper. Second RL model does not have action erase. It cannot modify the replication count of a block.

We compared our RL models against two possible baselines.

\subsection{Efficiency of RL models} \label{subsec:general_res}

\begin{figure*}[h] \label{fig:general_res_plot}
\centering
\includegraphics[width=1.0\textwidth]{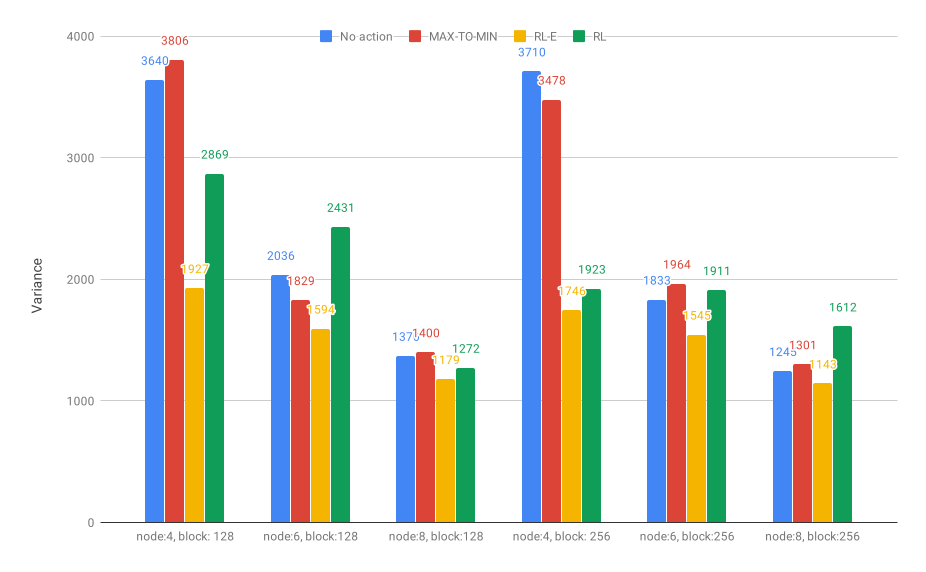}
\caption{Performance evaluation of RL models and comparable baselines.}
\end{figure*}

We evaluated RL model on various systems. The average variance of block read requests among nodes over timestep are shown on the figure \ref{fig:general_res_plot}. As expected, we can see that RL model(RL-E) consistently performs better than baselines. Notably, RL models are able to learn good strategies directly from iterations without any prior knowledge of the system.

However, performance gap between baselines and RL models gets small as the number of nodes increases. We think this is because the action space have size of $O(M^2)$ and gets too large for RL model to learn well.

\subsection{Convergence of RL Model} \label{subsec:convergence}

\begin{figure}[h] \label{fig:learning}
\centering
\includegraphics[width=0.45\textwidth]{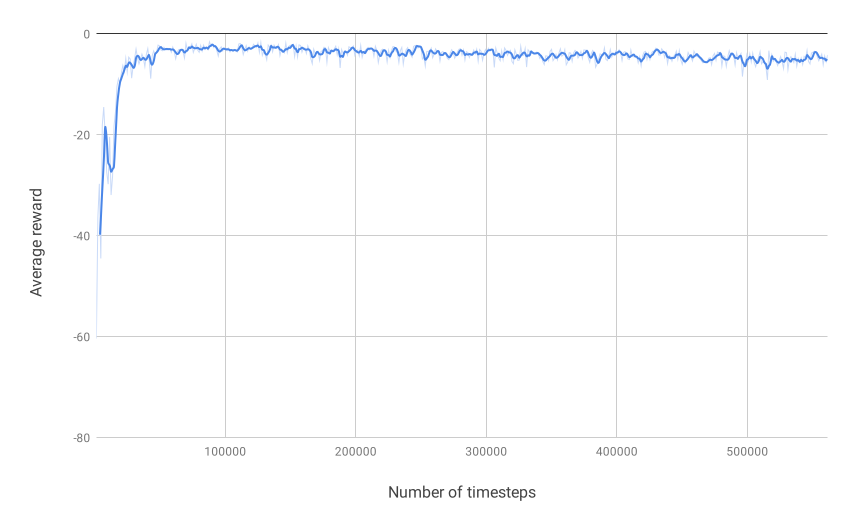}
\caption{Learning Curve showing that the model improves the total rewards toward a convergence. a line with shallow color is actual values, a line with deep blue color is trendline.}
\end{figure}

Figure \ref{fig:learning} plots the average rewards at each scenario. Rewards increase with timesteps as RL policy improves. We see the improvement over timestep as expected, and that the RL model seems to converge toward specific policy(average rewards fixed after 20k steps and more).

\begin{figure}[h] \label{fig:entropy}
\centering
\includegraphics[width=0.45\textwidth]{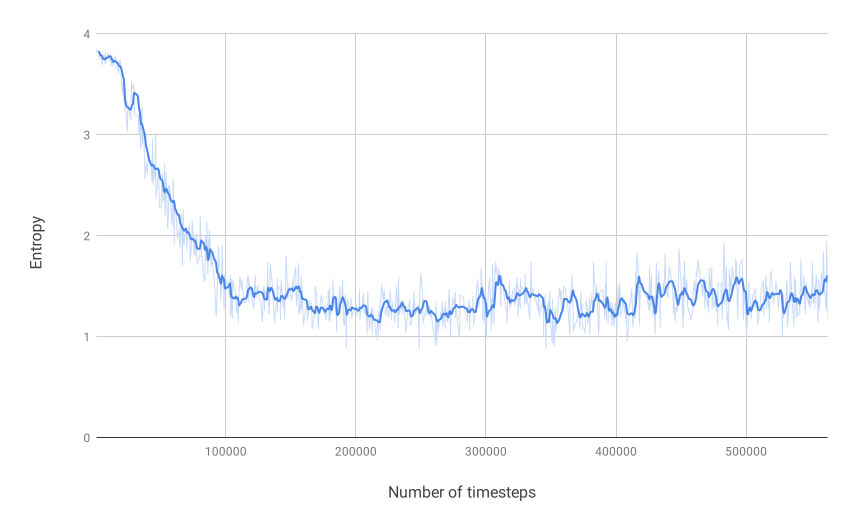}
\caption{Entropy of action distribution decreases. a line with shallow color is actual values, a line with deep blue color is trendline.}
\end{figure}

Recall that the policy gives probability distribution over actions. We can think that lower entropy assures the model converges to a specific policy. Figure \ref{fig:entropy} shows our RL model converges well.

\section{Conclusion} \label{sec:conclusion}
In this paper, we examined that it is feasible to apply Reinforcement Learning to system domain problems. Our experiments show that the RL model is comparable, even outperform than other heuristics for block management problem. However, our experiments are limited in terms of scalability and fidelity. Even though our formulation is not very practical, applying Reinforcement Learning to system domain could offer good alternatives to existing heuristics.

\bibliographystyle{plain}

\bibliography{main}

\end{document}